\documentclass{aa}

\makeatletter
\renewcommand*\aa@pageof{, page \thepage{} of \pageref*{LastPage}}
\makeatother

\usepackage{graphicx}
\usepackage{txfonts}
\usepackage[colorlinks=true,linkcolor=blue]{hyperref}
\hypersetup{citecolor=blue, urlcolor=blue}
\usepackage{amsmath}	
\usepackage{subfig}
\usepackage[normalem]{ulem}
\usepackage[separate-uncertainty=true]{siunitx}
\usepackage{dcolumn}
\usepackage{arydshln}    
\usepackage{tablefootnote}
\usepackage[dvipsnames]{xcolor}
\usepackage{orcidlink}
\usepackage{makecell}
\usepackage[flushleft]{threeparttable}
\usepackage{soul}

\newcommand{\g}{G288.8--6.3}
\newcommand{\rd}{\textit{RadialDisk}}
\newcommand{\rg}{\textit{RadialGaussian}}
\newcommand{\pl}{\textit{PowerLaw}}
\newcommand{\lp}{\textit{LogParabola}}

\DeclareSIUnit\erg{erg}
\DeclareSIUnit\SIsigma{\ensuremath{\sigma}}
\DeclareSIUnit\parsec{pc}
\DeclareSIUnit\photon{ph}

\newcolumntype{d}[1]{D{.}{.}{#1}} 

\begin{document}

   \title{Gamma-ray detection of newly discovered Ancora SNR:\\ G288.8--6.3}


   \author{Christopher Burger-Scheidlin
          \orcidlink{0000-0002-7239-2248}\inst{1,2}\thanks{Email: \href{mailto:cburger@cp.dias.ie}{cburger@cp.dias.ie}}
          \and
          Robert Brose \orcidlink{0000-0002-8312-6930}\inst{1,3}
          \and
          Jonathan Mackey \orcidlink{0000-0002-5449-6131}\inst{1}\fnmsep\inst{2}
          \and
          Miroslav D. Filipovi\'c \orcidlink{0000-0002-4990-9288}\inst{4}
          \and \\ \hspace{0.25cm}
          Pranjupriya Goswami \orcidlink{0000-0001-5430-4374}\inst{5, 6}
          \and
          Enrique Mestre Guillen \orcidlink{0000-0003-3968-1782}\inst{7}
          \and
          Emma de O\~na Wilhelmi \orcidlink{0000-0002-5401-0744}\inst{8}
          \and
          Iurii Sushch \orcidlink{0000-0002-2814-1257}\inst{6, 9}
          }

   \institute{Dublin Institute for Advanced Studies, Astronomy \& Astrophysics Section, DIAS Dunsink Observatory, Dublin, D15 XR2R, Ireland
        \and
        School of Physics, University College Dublin, Belfield, Dublin, D04 V1W8, Ireland
        \and
        School of Physical Sciences and Centre for Astrophysics \& Relativity, Dublin City University, D09 W6Y4 Glasnevin, Ireland
        \and
        Western Sydney University, Locked Bag 1797, Penrith South DC, NSW 2751, Australia
        \and
        Université Paris Cité, CNRS, Astroparticule et Cosmologie, F-75013 Paris, France
        \and
        Centre for Space Research, North-West University, 2520 Potchefstroom, South Africa
        \and
        Institute of Space Sciences (ICE, CSIC), Campus UAB, Carrer de Can Magrans s/n, 08193 Barcelona, Spain
        \and
        Deutsches Elektronen-Synchrotron DESY, Platanenallee 6, 15738 Zeuthen, Germany
        \and
        Astronomical Observatory of Ivan Franko National University of Lviv, Kyryla i Methodia 8, 79005 Lviv, Ukraine
        }

   \date{Received 23 Oct 2023; accepted 23 Jan 2024}

    \authorrunning{C. Burger-Scheidlin et al.}

\abstract{The supernova remnant (SNR) \g\ was recently discovered as a faint radio shell at large Galactic latitude using observations with the \textit{Australian Square Kilometre Array Pathfinder} (\textit{ASKAP}) in the Evolutionary Map of the Universe (EMU) survey.}
{Here, we make the first detailed investigation of the $\gamma$-ray emission from the \g\ region, aiming to characterise the high-energy emission in the GeV regime from the newly discovered SNR, dubbed Ancora.}
{\SI{15}{} years of \textit{Fermi}--Large Area Telescope (LAT) data were analysed at energies between \SI{400}{\MeV} and \SI{1}{\TeV} and the excess seen in the region was modelled using different spatial and spectral models.}
{We detect spatially extended $\gamma$-ray emission coinciding with the radio SNR, with detection significance up to \SI{8.8}{\SIsigma}. A radial disk spatial model in combination with a power-law spectral model with an energy flux of \SI{4.80\pm0.91e-06}{\MeV\per\square\cm\per\s}, with the spectrum extending up to around \SI{5}{\GeV} was found to be the preferred model. Morphologically, hotspots seen above \SI{1}{\GeV} are well-correlated with the bright western part of the radio shell. The emission is more likely to be of leptonic origin given the estimated gas density in the region and the estimated distance and age of the SNR, but a hadronic scenario cannot be ruled out.}
{Ancora is the seventh confirmed SNR detected at high Galactic latitude with \textit{Fermi}--LAT. The study of this new population of remnants can help gain insight into the evolutionary aspects of SNRs and their properties, and further advance efforts of constraining the physics of particle diffusion and escape from SNRs into the Galaxy.}

               
\maketitle




\section{Introduction}

After the detection of cosmic rays (CRs) through observations of ionising radiation on multiple balloon flights by Victor Hess~\citep{Hess_1912}, their spectrum was subsequently measured by many experiments \citep[e.g.,][]{Tanabashi_2018, Filipovic_Book_2021}. The current consensus is that cosmic protons with energies up to \SI{3e15}{eV} have Galactic origin~\citep{Aloisio_2012, Globus_2015}.
Supernova remnants (SNRs) have been traditionally considered to be prime candidates for the acceleration of these particles to high energies and were first suggested as sources by~\cite{Baade_and_Zwicky_1934}.
The number of currently detected and confirmed SNRs lies close to three hundred~\citep{Green_2017, Green_2019}, the majority of them being first detected at radio wavelengths.
Around ten percent are subsequently also detected at $\gamma$-ray energies~\citep{Acero_2016}, only $\approx\!\!10$ remnants where non-thermal X-ray synchrotron emission was seen~\citep{2012A&ARv..20...49V}, and even fewer have been detected at very-high-energy (VHE) $\gamma$-rays~\citep[e.g.,][]{HESS_2018}.

Recently the population of high-latitude supernova remnants has received some attention, with \cite{Ackermann_2018} finding SNR candidates amongst the analysed sources, as well as the detection of G296.5+10.0 and G166.0+4.3~\citep{Araya_2013}, G150+4.5~\citep{Devin_2020}, G17.8+16.7~\citep{Araya_2021} and Calvera SNR~(\citealp{Arias_2022} and \citealp{Araya_2022}) emitting in this high-energy $\gamma$-ray regime, amongst others.

In their first published SNR catalogue the Fermi collaboration classify 30 sources as likely \SI{}{\giga\electronvolt} SNRs, as well as four marginal associations and 245 flux upper limits~\citep{Acero_2016}.
Although instruments have improved significantly since early high-energy surveys for SNRs with, e.g.~\textit{EGRET}~\citep{Sturner_1995, Esposito_1996}, the limited spatial resolution and photon statistics in $\gamma$-rays make new discoveries challenging.
Most SNRs are therefore first identified by their distinctive non-thermal radio spectrum and shell-like morphology.
Observations at high energies allow for a much more detailed understanding of the physical phenomena taking place at the sources. They potentially permit distinguishing between leptonic and hadronic emission scenarios~\citep[e.g.,][]{Aharonian_2007}, as well as being able to constrain maximum particle energies \citep[e.g.,][]{VERITAS_2020}.
Morphological differences between wavelength bands can also give insight into where particle acceleration is taking place and how CRs escape from SNRs into the Galaxy.
Furthermore, increasing the number of SNRs detected at high energies enables comparisons between these different sources, enhancing our understanding of the population class to find common spectral and spatial features as well as identifying unexpected properties.

Here we report the detection of extended $\gamma$-ray emission spatially coincident with the shell of a newly discovered radio SNR.
Ancora SNR\footnote{The location of the remnant is close to where the anchor (\textit{Latin: ancora}) of \textit{Argo Navis} would be in the constellation of \textit{Carina}. Details can be found in appendix \ref{appendix:ancora}.} (also known as \g) at a high Galactic latitude of \SI{-6.3}{\degree} was first detected by \cite{Filipovic_2023} through multi-frequency analysis of \textit{Australian Square Kilometre Array Pathfinder (ASKAP)} data. The authors reported a low radio-surface-brightness SNR with a shell-like structure extending to a diameter of $\SI{1.8}{\degree}\times\SI{1.6}{\degree}$. 
They estimate the distance at $\approx$\SI{1.3}{\kilo\parsec}, implying an offset of \SI{140}{\parsec} above the Galactic plane, and the age of the remnant, based on the surface brightness, at $>$\SI{13}{\kilo yr}.

The detection in radio continuum wavebands, with strong hints of non-thermal emission, sparked the efforts to to analyse the region at $\gamma$-ray energies using the \textit{Fermi}--LAT instrument.
In the following sections, we will show how we performed the analysis of LAT data and which models were considered to fit the excess emission (section~\ref{sec:data-analysis}), which major results we were able to obtain from the best-fit models and how they relate to other instruments (section~\ref{sec:results}).
In section~\ref{sec:discussion} we discuss our findings in the context of the other high-latitude SNRs and the potential for further detections at higher and lower energies. Finally, we present our conclusions in section~\ref{sec:conclusions}.

\section{Fermi-LAT data analysis}
\label{sec:data-analysis}

\subsection{Data reduction}
We used 15 years (August 2008 -- July 2023, correspondingly \num{239557417}\:--\:\num{712342367} \textit{Fermi} mission elapsed time (MET)) of \textit{Fermi}--LAT Pass 8 data using the \texttt{P8R3\_SOURCE} class events~\citep{Atwood_2013}. The selected data was within a radius of \SI{12}{\degree} in the region of interest (ROI) around the centre of the detected radio SNR at Galactic position GLON/GLAT~$ =\SI{288.8}{\degree}$/$\SI{-6.3}{\degree}$ (corresponding to $\text{R.A./Dec} = \SI{157.488}{\degree}$/$\SI{-65.214}{\degree}$) in a range between \SI{400}{\mega\electronvolt} and \SI{1}{\tera\electronvolt} energies. We analysed only data above \SI{400}{\MeV} to decrease the uncertainties due to the poorer angular resolution, given the large size of the SNR. Additionally, cuts on the PSF class were made by employing \texttt{evtype = 56}, discarding the lowest-quality quartile data in terms of event direction reconstruction (i.e., \texttt{PSF0} class events).

The data were analysed using the Python package \textit{Fermipy}~(v1.1.6, \citealp{Fermi_2017}) and the \textit{Fermitools} software package (v2.2.0\footnote{\href{https://github.com/fermi-lat/Fermitools-conda}{github.com/fermi-lat/Fermitools-conda}}) following the standard procedure of the binned maximum-likelihood analysis technique together with the \texttt{P8R3\_V3} instrument response functions. The recommended filters ($\texttt{DATA\_QUAL}>0 \; \texttt{\&\&} \; \texttt{LAT\_CONFIG}==1$) were used for data reduction, as well as zenith angle cuts of \SI{90}{\degree} applied to reduce contamination from the direction of Earth's atmosphere.
Regarding spatial binning, a pixel size of \SI{0.1}{\degree} with eight logarithmic energy bins per decade was applied. The underlying source catalogue used was the fourth Fermi catalogue~\citep{Abdollahi_2020}, specifically the incremental Data Release~3 version 4FGL-DR3~\citep{Abdollahi_2022}, modelling sources that were lying up to \SI{3}{\degree} outside the ROI. To account for diffuse emission, we applied the Galactic diffuse emission model (\texttt{gll\_iem\_v07.fits}) with an isotropic component corresponding to the chosen event class (\texttt{iso\_P8R3\_SOURCE\_V3\_v1.txt}).

\begin{table*}
	\centering
	\caption{List of different models and their relative log-likelihood values compared to the base model, with $\mathcal{L}_0 = 435708.381$, and $\Delta ln(\mathcal{L}) = ln(\mathcal{L}_\text{i}) - ln(\mathcal{L}_0)$, and difference in model parameters compared to the base model ($\Delta k = k_\text{i} - k_\text{0}$). The two best-fit models, both presented in this work, are marked in bold font.}
	\label{tab:model_comparison}
    \begin{threeparttable} 
	\begin{tabular}{cccllccllll} 
    \\
    \hline\hline
    ~ & Model N$^\text{o}$ & J1028 incl. & Spatial model & Spectral model & $\Delta ln(\mathcal{L})$ & $\Delta k$ & $\Delta$AIC & ~ & ~ & ~ \\ \hline
    ~ & \texttt{0} & Y & --- & --- & 0 & \,\,\,0 & \,\,\,\,\,\,\,0 & ~ & ~ & ~  \\
    ~ & \texttt{1} & N & --- & --- & $-$24.87 & $-$5 & \,\,\,\,39.74 & ~ & ~ & ~ \\

    \hdashline[0.75pt/1pt]

    ~ & \textbf{\texttt{2}} & \textbf{Y} & \textbf{\rd} & \textbf{\pl} & \,\,\,\textbf{16.50} & \,\,\,\textbf{5} & \textbf{$-$22.99} & ~ & ~ & ~ \\
    ~ & \textbf{\texttt{3}} & \textbf{N} & \textbf{\rd} & \textbf{\pl} & \,\,\,\,\,\,\textbf{9.54} & \,\,\,\textbf{0} & \textbf{$-$21.08} & ~ & ~ & ~ \\
    ~ & \texttt{4} & N & \rd & \lp & \,\,\,\,\,\,9.80 & \,\,\,1 & $-$19.59 & ~ & ~ & ~ \\

    \hdashline[0.75pt/1pt]

    ~ & \texttt{5} & N & Radio template & \pl & \,\,\,\,\,\,7.76 & $-$1 & $-$19.51 & ~ & ~ & ~ \\

    \hline\hline
   
	\end{tabular}

    \end{threeparttable}
\end{table*}

\subsection{Morphological and spectral analysis}
\label{sec:spectral-analysis}
Excess from SNR \g\ has only recently been discovered at radio wavelengths (\SI{954}{\mega\hertz}) with the \textit{Australian Square Kilometre Array Pathfinder (ASKAP)} by~\cite{Filipovic_2023}. The authors found a spatial extension of $\SI{1.8}{\degree}\times\SI{1.6}{\degree}$ in diameter of this faint, low-surface brightness radio source.
When searching for excess in $\gamma$-rays we analysed the energy range between \SI{400}{\mega\electronvolt} to \SI{1}{\tera\electronvolt}, and later made cuts on the energy range to confirm the extension, find morphological features and excess at higher energies where the point-spread-function (PSF) of the \textit{Fermi}--LAT instrument is much better compared to lower energies\footnote{\href{https://www.slac.stanford.edu/exp/glast/groups/canda/lat_Performance.htm}{slac.stanford.edu/exp/glast/groups/canda/lat\_Performance.htm}}.
Except for one source, namely 4FGL\,J1028.7-6431c (henceforth referred to simply as J1028), all sources in the 4FGL-DR3 catalogue around the target position were at distances larger than \SI{1.3}{\degree} from the centre of the radio source.
J1028 is located around \SI{0.7}{\degree} from the centre of the SNR and is thus roughly overlapping with the northern part of its shell. It is catalogued as a slightly elongated source with a \SI{68}{\percent} containment radius of $\sim \SI{0.12}{\degree}$, and a log-parabola spectral model of the form

\begin{equation}
\centering
\label{eq:logparabola}
    \frac{\text{dN}}{\text{dE}} = \text{N}_0\left(\frac{\text{E}}{\text{E}_0}\right)^{- \alpha - \beta \log(\text{E}/\text{E}_0)}\text{ , } 
\end{equation}

\noindent with an energy flux $\text{N}_0 = \SI{1.5\pm0.3e-12}{\per\mega\electronvolt\per\centi\metre\squared\per\second}$, $\alpha=2.6\pm0.3$ and $\beta=0.4\pm0.2$, and a pivot energy of $\text{E}_0=\SI{720}{\mega\electronvolt}$. It has a detection significance of \SI{5.6}{\SIsigma}. An association with PMN~J1028-6441 was claimed earlier~\citep{Abdollahi_2022}.

To test different models, J1028 was removed from some of the analyses for comparison of the goodness of fit.
Spectral parameters of the sources within the \SI{3}{\degree} were left free during fitting, as well as normalisation factors of sources above a test statistic,\footnote{TS is defined as twice the difference between log-likelihoods of source plus background ($\mathcal{L}_1$) and only background as a null hypothesis ($\mathcal{L}_0$):  $\text{TS} =2\,\text{ln}(\mathcal{L}_1/\mathcal{L}_0$).} $\text{TS}=10$ within the ROI. The parameters of all other sources were kept fixed to their 4FGL-DR3 catalogue values.
We applied the energy dispersion correction for all sources apart from the isotropic diffuse emission background model.
We performed three initial optimisations of the sources using the \texttt{optimize()} method which attempts a preliminary fit to ensure parameters are close to their maximum likelihood values, and then an initial fit using the \texttt{fit()} method with the \texttt{newminuit} optimiser. Afterwards, we removed sources with $\text{TS}<1$. Further fitting showed that no additional sources with $\text{TS}<1$ appeared.
Subsequently, we added different extended spatial models (\rd\ and \rg) using the \texttt{add\_source()} method at the known radio source position (GLON/GLAT = \SI{288.8}{\degree}/\SI{-6.3}{\degree}) with their default extension parameters, and refitted the region.
Then, the \texttt{extension()} method provided in \textit{Fermipy} was applied to determine the best-fit radius iterating between the extension parameters of \SI{0.4}{\degree} to \SI{1.5}{\degree} in 21 logarithmic bins by performing a likelihood profile scan that maximises the model likelihood. The resulting best-fit extension was then added to the model, replacing the default value, and the whole region was refit again. The initial position from the radio detection was left unchanged to see spatial overlap with the radio source and determine the high-energy flux coming from that source region at this position, only varying the radius.
We checked the source extension for a radial disk model and radial, symmetric two-dimensional (2D) Gaussian, as well as fitting a template of the smoothed radio template, using a power-law (PL) as well as a log-parabola (LP, equation~\ref{eq:logparabola}) spectral model.

The power-law spectral model used in \textit{Fermipy} is of the form

\begin{equation}
    \frac{\text{dN}}{\text{dE}} = \text{N}_0 \left(\frac{\text{E}}{\text{E}_0}\right)^{-\Gamma} \,\,\,,
\end{equation}

\noindent with N$_0$ the normalisation/prefactor, $\Gamma$ the spectral index, and E$_0$ the scaling energy.
During the fitting the spectral parameters of the sources within \SI{3}{\degree} were left to vary. 

We did not optimise the source position as the aim of this study was to test for a $\gamma$-ray source at the radio position. The radial size of the model was, however, left to vary as the emission may be larger or smaller than the radio extension.

While performing the analysis it became clear that the \rg\ model was systematically overestimating the extension of the source by associating events previously modelled by the background or other sources to it and subsequently reaching very high extensions ($\sim\SI{1.5}{\degree}$). These extension results additionally varied strongly with different energy cuts. Despite the overall likelihood of the models seeming favourable, the authors agreed that the \rg\ model was not well-behaved and unsuitable for the underlying purposes. Thus, the analysis efforts were focused on the \rd\ model and \rg\ results are not shown in this work.

\begin{figure*}
    \captionsetup[subfigure]{labelformat=empty}
    \subfloat[\centering]{{\includegraphics[width=0.96\columnwidth, trim={0.7cm 0cm 3.5cm 0cm}, clip]{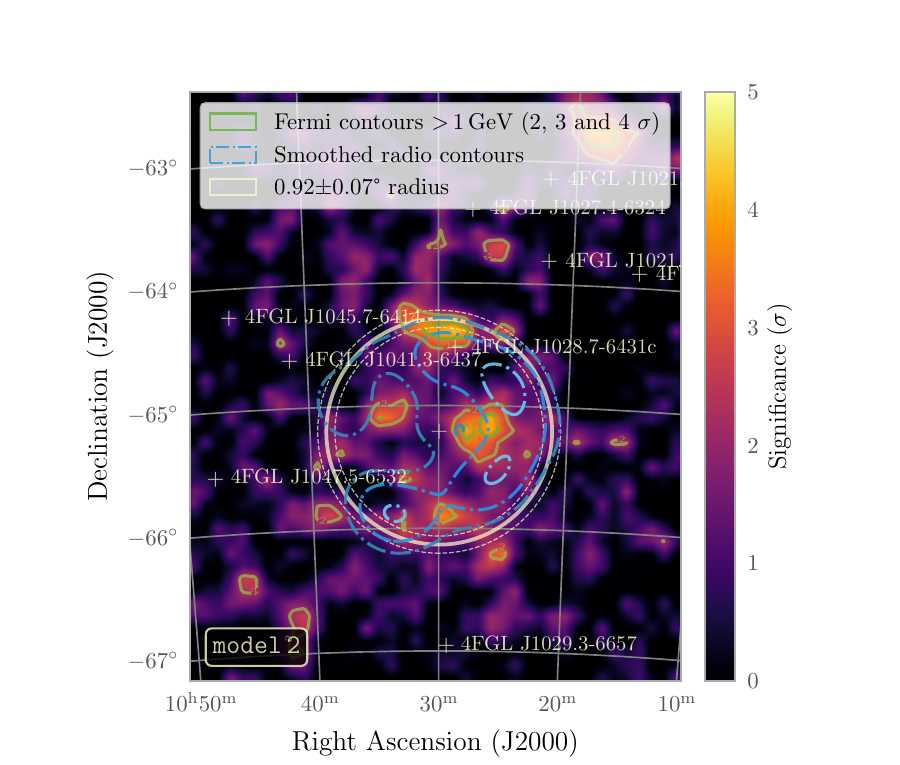}}}%
    \qquad
    \subfloat[\centering]{{\includegraphics[width=\columnwidth, trim={2.2cm 0cm 1.5cm 0cm}, clip]{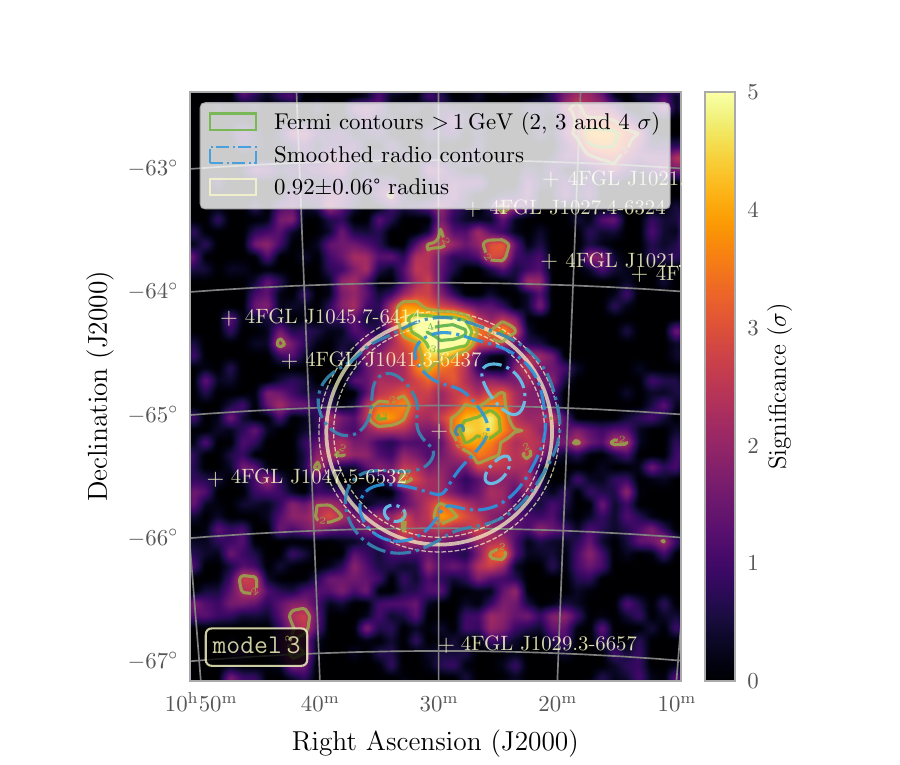}}}%
    \vspace{-0.4cm}
    \caption{Significance maps of the \g\ region as seen with \textit{Fermi}--LAT after fitting with a \rd\ spatial model and a \pl\ spectral model with (left, \texttt{model\,2}) and without (right, \texttt{model\,3}) modelling J1028 in the energy range of $\SI{400}{\MeV} - \SI{1}{\TeV}$. The circles with radius \SI{0.92}{\degree} show the disk radius obtained from the respective fits. The plots show the significance map after subtracting all the fitted sources, except for the SNR. The maps are overlaid with Fermi contours for significance values above \SI{1}{\giga\electronvolt}, and the smoothed radio contours from the ASKAP instrument at \SI{954}{\mega\hertz}.
    }
    \label{fig:sign_map}
\end{figure*}

\begin{table*}
	\centering
	\caption{Best-fit parameters of the \rd\ model with and without modelling J1028 for Ancora SNR, as well as the results for the spatial template derived from the radio signal from the object.}
	\label{tab:best_fit_model_params}
\begin{threeparttable}
\begin{tabular}{lllll} \\
\hline\hline
Parameter & Unit & \multicolumn{3}{c}{Value} \\ \hline

Position \\
\hspace{0.6cm} R.A. / Dec & deg / deg & \multicolumn{3}{c}{157.488 / $-65.214$} \\
\hspace{0.6cm} GLON / GLAT & deg / deg & \multicolumn{3}{c}{288.8 / $-6.3$\,\,\,}\\
\hdashline[0.75pt/1pt]

Model N$^\text{o}$ &  & \texttt{2} & \texttt{3} & \texttt{5} \\
J1028 incl. &  & Y & N & N \\
\hdashline[0.75pt/1pt]

Spatial model &  & \rd & \rd & \textit{SpatialTemplate} \\
Spectral model &  & \pl & \pl & \pl \\
\hdashline[0.75pt/1pt]

TS & --- & 44.74 & 77.14 & 70.98 \\
N$^\text{o}$ of predicted photons & --- & 978 & 1331 & 1174 \\
Photon flux & \SI{}{\photon\per\square\centi\metre\per\second} & \SI{2.29\pm0.45e-09}{} & \SI{3.14\pm0.41e-09}{} & \SI{2.74\pm0.37e-09}{}\\
Energy flux & \SI{}{\mega\electronvolt\per\centi\metre\squared\per\second} & \SI{4.29\pm1.03e-06}{} & \SI{4.80\pm0.91e-06}{} & \SI{3.62\pm0.68e-06}{} \\
\hspace{0.6cm} > \SI{1}{\GeV} (to \SI{316}{\GeV}) & \SI{}{\mega\electronvolt\per\centi\metre\squared\per\second} & \SI{3.08\pm0.83e-6}{} & \SI{3.29\pm0.78e-6}{} & \SI{2.13\pm0.66e-6}{} \\
\hdashline[0.75pt/1pt]

Spectral parameters \\
\hspace{0.6cm} N$_0$ & \SI{}{\per\mega\electronvolt\per\centi\metre\squared\per\second} & \SI{9.17\pm1.81e-13}{} & \SI{1.23\pm0.16e-12}{} & \SI{1.15\pm0.15e-02}{} \\
\hspace{0.6cm} $\Gamma$ & --- & $2.21\pm0.12$ & $2.32\pm0.11$ & $2.41\pm0.13$ \\
\hspace{0.6cm} E$_0$ & \SI{}{MeV} & $1000$\tnote{*} & $1000$\tnote{*} & $1000$\tnote{*} \\
\hdashline[0.75pt/1pt]

Spatial parameters\\
\hspace{0.6cm} Extension & deg & $0.92\pm0.07$ & $0.92\pm0.06$ & --- \\
\hspace{0.6cm} TS$_\text{ext}$\tnote{\dag} & --- & 33.92 & 52.56 & --- \\

\hline\hline
 
 \end{tabular}

\begin{tablenotes}
        \scriptsize
        \item[*] Parameter fixed
        \item[\dag] Test statistic for the extension hypothesis against the null hypothesis of a point-like source
        
    \end{tablenotes}

 \end{threeparttable}
\end{table*}

\section{Results}
\label{sec:results}

\subsection{Comparing models}

In this study, we tested different spatial models as well as two different spectral models for the best-fit model. A list of different iterations of the spatial models is shown in Table~\ref{tab:model_comparison}.
Models were numbered starting with the base model (\texttt{model\,0}) up to \texttt{model\,5}. The second column indicates whether the model for J1028 was included in the fitting (Y) or not (N). 
The relative log-likelihoods ln($\mathcal{L}_\text{i}$) compared to the base model ln($\mathcal{L}_0$) are given, as well as the relative number of free fitting parameters $k_\text{i}$. Positive relative log-likelihood values indicate higher likelihood, and thus a better qualitative fit to the underlying data. Lower values indicate a worse fit. Additionally, we show the value of the Akaike Information Criterion (AIC, \citealp{Akaike_1974}), a qualitative measure that indicates which of a number of different models is preferred over the others, taking into account the difference in free parameters. The $\text{AIC} = 2 k - 2\,\text{ln(}\mathcal{L})$ is presented in Table~\ref{tab:model_comparison} as the difference between AIC values of different models, $\Delta$AIC. Designated models are again compared to the base model ($\Delta \text{AIC} = \text{AIC}_\text{i} - \text{AIC}_0$).

Table~\ref{tab:model_comparison} first shows the base model, the unchanged 4FGL-DR3 catalogue results (\texttt{model\,0}) and then the identical model except that J1028 was removed from the modelling (\texttt{model~1}), as indicated by the second column, `J1028 incl.'. To model the excess, for \texttt{models~2}, \texttt{3}, and \texttt{4}, a \rd\ spatial model was applied, including or excluding J1028 in the modelling\footnote{When adding the extended source like in fit \texttt{models\,2-4}, the catalogue sources as well as various background model normalisations can change as the overall region is optimised for maximising the log-likelihood value. Therefore the \texttt{model\,0} may differ from \texttt{model\,2} when removing the extended source after fitting, and the same applies for \texttt{model\,1} and \texttt{models\,3-4}.}.

Finally, we also show the performance of the model derived from the radio template (\texttt{model\,5}).

\subsection{Best-fit models and spatial results}

The $\Delta$AIC value for \texttt{model\,2} is the lowest, indicating that it is preferred over the other models (highlighted in bold font in Table~\ref{tab:model_comparison}). As this model includes the modelling of J1028, which lies within \SI{0.7}{\degree} of the centre of the SNR, it is quite possible that this excess is part of Ancora and a combination of the employed disk model together with the model for J1028 are the best description of the SNR.

We note that \texttt{model\,3} also obtained values close to the best model (also highlighted in bold font in Table~\ref{tab:model_comparison}), although some unmodelled excess can be seen in the northern part of the shell. It is thus not entirely clear whether J1028 is actually associated with Ancora, or is an unrelated source overlapping with the SNR. Spectral analysis (see below) shows that Ancora and J1028 have different spectral shape, although given the limited photon counts this difference may not be significant.

The results of \texttt{model\,2} and \texttt{3} (the \rd\ spatial model including and excluding the J1028 model from the source list, and a \pl\ spectral model) are shown in Fig.~\ref{fig:sign_map}. The figure shows the resulting significance maps within a field of view (FoV) of approximately $\SI{4.0}{\degree}\times\SI{4.5}{\degree}$ for an energy range of \SI{400}{\mega\electronvolt} -- \SI{1}{\tera\electronvolt}, with overlays (2, 3 and \SI{4}{\SIsigma}) of the resulting peaks above \SI{1}{\giga\electronvolt}, as well as smoothed radio contours \citep{Filipovic_2023} and the best-fit \SI{0.92}{\degree} radius for these model around the source position for easier orientation.
The significance maps both shows an extended peak in $\gamma$-rays that coincides with the radio contours.
Overall, the morphology is not identical: the high-energy emission is slightly more extended than the radio, and the clear shell morphology of the radio map is not seen in $\gamma$-rays.
Flux contours for photons with energy above \SI{1}{\giga\electronvolt} show much better correspondence with radio emission, tracing the western and northern parts of the shell in radio.
To check the reliability of the extension fitting, an analysis for energies above \SI{800}{\MeV} was performed which resulted in an extension of \SI{0.91}{\degree}, consistent with extensions found at lower energies.

\subsection{Spectral results}
The spectral energy distribution (SED) of the \g\ region described with the \rd\ model without modelling J1028 (\texttt{model\,3}) extends up to \SI{5}{\GeV} and fits a \pl\ spectral model with an energy flux of \SI{4.80\pm0.91e-06}{\MeV\per\square\cm\per\s} (photon flux of \SI{3.14\pm0.41e-09}{\photon\per\square\cm\per\s}), and spectral index of $\Gamma=\SI{2.32}{} \pm \SI{0.11}{}$, accounting only for statistical errors. The total significance of the source was estimated to be $\text{TS}=\SI{77}{}$, or an equivalent of $\sim\SI{8.8}{\SIsigma}$. The model improves the fit compared to the original catalogue with an increase of 9.54 in the log-likelihood value while maintaining the overall number of free parameters. 

\begin{figure*}
    \centering
    \captionsetup[subfigure]{labelformat=empty}
    \subfloat[\centering ]{{\includegraphics[width=0.91\columnwidth, trim={0cm 0cm 0.cm 0cm}, clip]{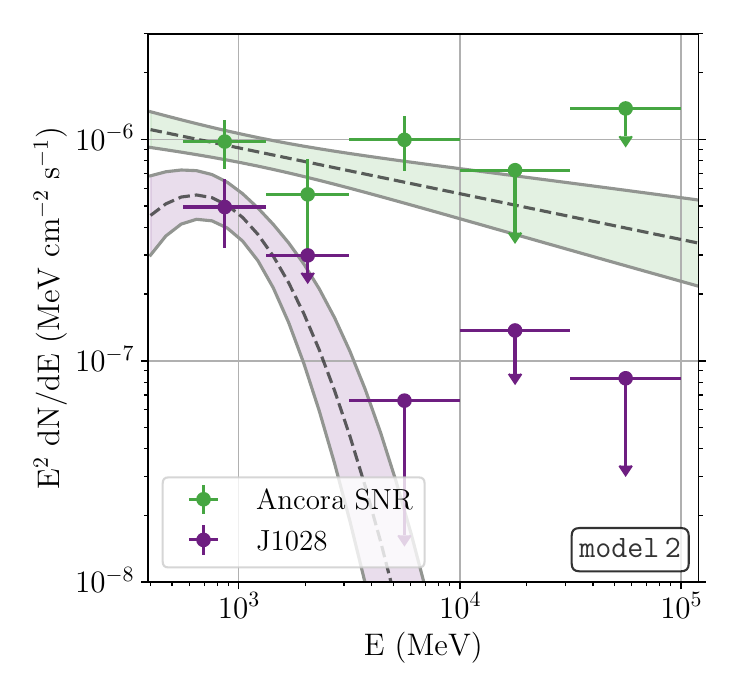}}}%
    \qquad
    \qquad
    \subfloat[\centering ]{{\includegraphics[width=0.91\columnwidth, trim={0cm 0.2cm 0.cm 0cm}, clip]{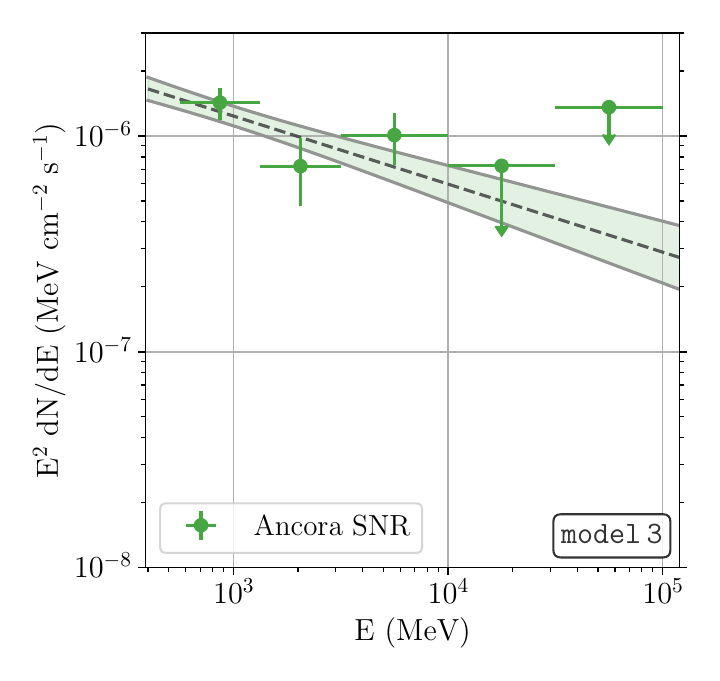}}}%
    \vspace{-0.6cm}
    \caption{Spectral energy distribution (SED) of the \g\ region using a \rd\ spatial model and a \pl\ with (left, \texttt{model\,2}) and without (right, \texttt{model\,3}) modelling J1028. Additionally, the spectral models as well as their \SI{1}{\SIsigma} uncertainty bands are shown in the respective colours of the spectral points.}
    \label{fig:sed}
\end{figure*}

When examining the results for \texttt{model\,2} (including modelling for J1028) we find that the TS of the extended disk decreases to 45 (significance of \SI{6.7}{\SIsigma}), with most of this reduction corresponding to the addition of flux to J1028. 
For this model we find that the disk model also extends up to \SI{5}{\GeV}, and fits a \pl\ spectral model with an energy flux of \SI{3.08\pm0.83e-06}{\MeV\per\square\cm\per\s} (photon flux of \SI{2.29\pm0.45e-09}{\photon\per\square\cm\per\s}), and spectral index of $\Gamma=\SI{2.21}{} \pm \SI{0.12}{}$. For J1028 we find an energy flux of \SI{7.49\pm2.16e-07}{\MeV\per\square\cm\per\s} (photon flux of \SI{9.68\pm3.08e-10}{\photon\per\square\cm\per\s}) with a total TS of 19.08 (significance of \SI{4.4}{\SIsigma}). Combining the model found for the SNR with the model for J1028, the total energy flux coming from this region is thus $\sim\SI{5.04e-06}{\MeV\per\square\cm\per\s}$ with a TS of 63.82 (\SI{8.0}{\SIsigma}). This improves the overall likelihood by 16.50 and is the best-fit model of the region (\texttt{model~2}), sightly though possibly not significantly preferred over \texttt{model\,3}.
All values and a direct comparison between \texttt{models\,2},\texttt{\,3} and \texttt{5} can be found in Table~\ref{tab:best_fit_model_params}. The SEDs for \texttt{models 2} and \texttt{\,3} can be found in Fig.~\ref{fig:sed}.

\begin{figure*}
    \centering
    \captionsetup[subfigure]{labelformat=empty}
    \subfloat[\centering]{{\includegraphics[width=0.83\columnwidth, trim={1.2cm 0cm 3.4cm 0.8cm}, clip]{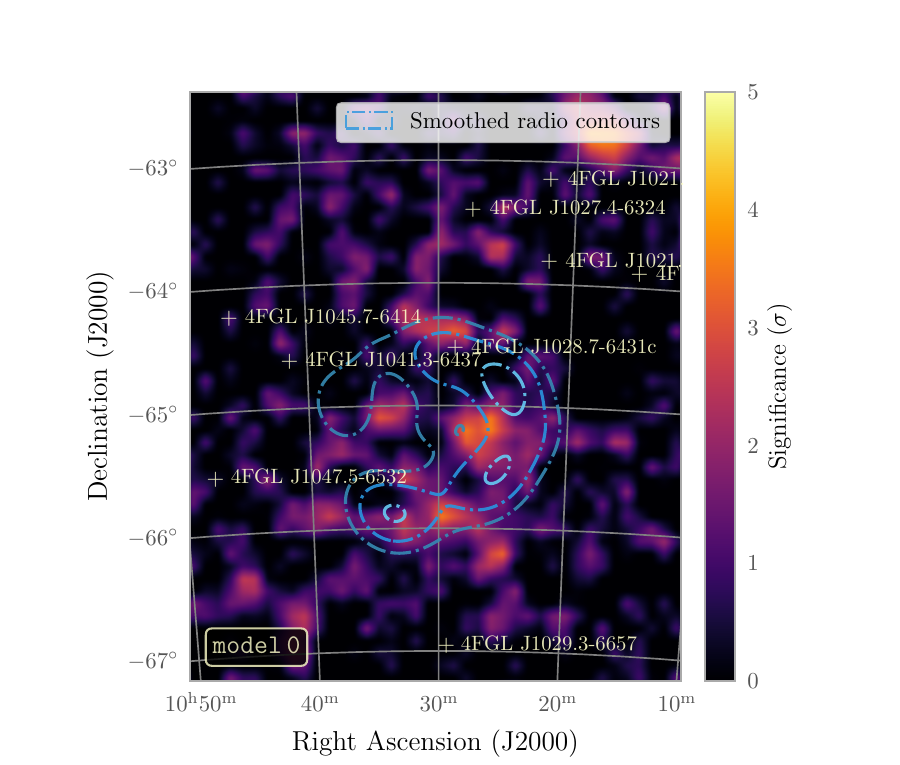}}}%
    \qquad
    \subfloat[\centering]{{\includegraphics[width=0.9\columnwidth, trim={2.2cm 0cm 1.5cm 0.8cm}, clip]{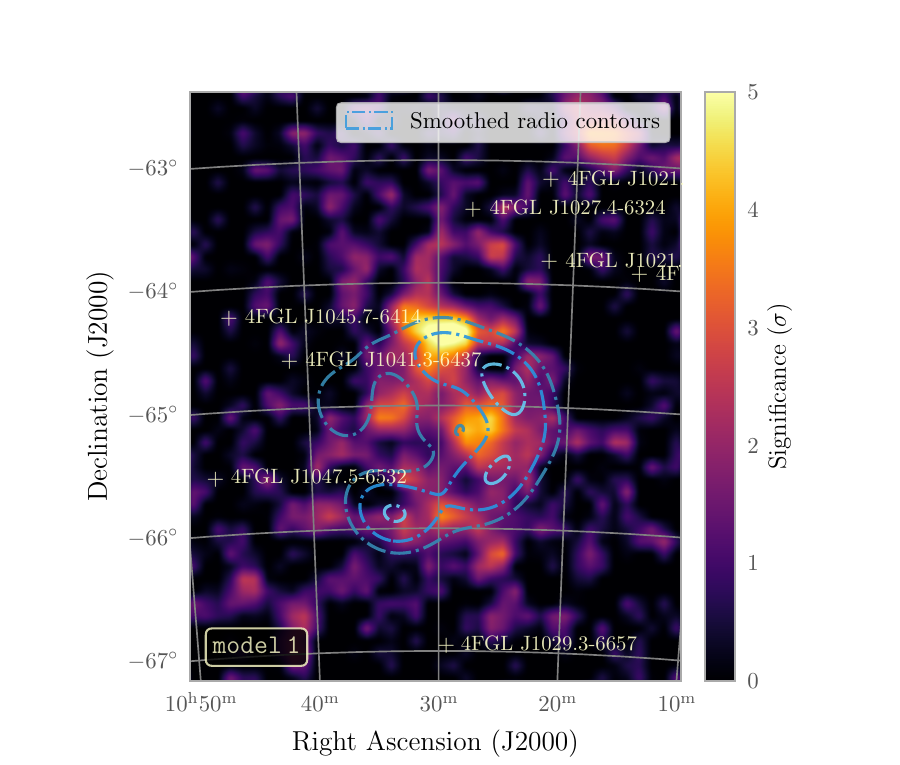}}}%
    \vspace{-1.4cm}
    \qquad
    \subfloat[\centering]{{\includegraphics[width=0.83\columnwidth, trim={1.2cm 0cm 3.4cm 0.8cm}, clip]{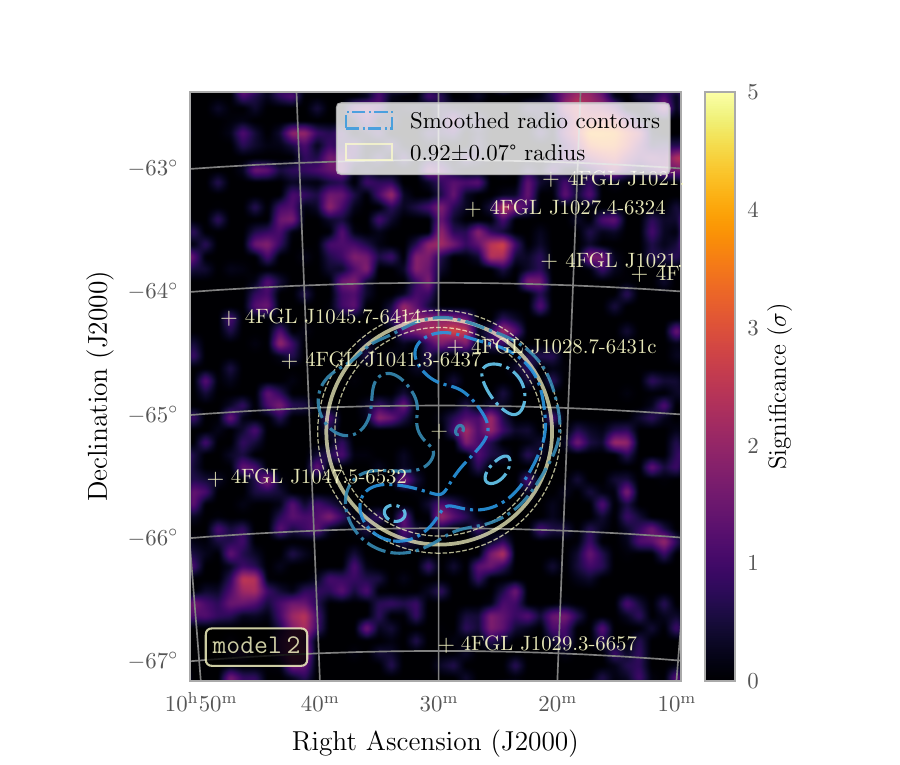}}}%
    \qquad
    \subfloat[\centering]{{\includegraphics[width=0.9\columnwidth, trim={2.2cm 0cm 1.5cm 0.8cm}, clip]{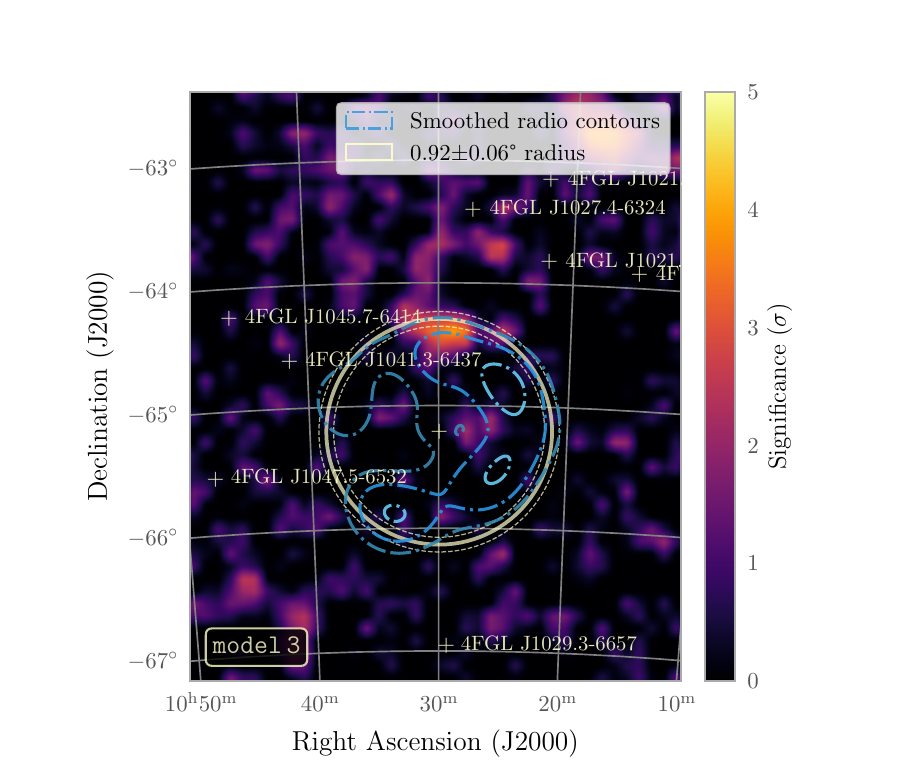}}}%
    \vspace{-0.4cm}
    \caption{\textit{Top row}: Residual maps of \texttt{model\,0} (top left, obtained directly from the 4FGL-DR3 catalogue) and \texttt{model\,1} (top right). \textit{Bottom row}: Residual maps for \texttt{model\,2} (bottom left), and \texttt{model\,3} (bottom right, position of J1028 marked just for reference), after adding a \rd\ spatial model and a \pl\ spectral model and refitting. All fits were in the energy range of $\SI{400}{\MeV} - \SI{1}{\TeV} $.
    Blue contours show smoothed radio emission. \texttt{Model\,0} and \texttt{1} show gamma-ray excess overlapping with this region. The circles with radius \SI{0.92}{\degree} in the bottom panels show the disk radius obtained from the fits of \texttt{model\,2} and \texttt{3}.
    Please note that the top and bottom panels represent different fits, as the region is refit after adding the extended SNR model. The bottom panels represent the exact fit shown in Fig.~\ref{fig:sign_map}, except for the removal of the extended model.
    }

    \label{fig:signmap_compare}
\end{figure*}

Furthermore, the residual map for \texttt{models~0} and \texttt{1} are shown in Fig.~\ref{fig:signmap_compare}~(top left and top right, respectively), showing the relatively high level of extended residual flux that cannot be assigned to any of the sources in the 4FGL-DR3 catalogue.
Fig.~\ref{fig:signmap_compare}~(bottom left and bottom right, respectively) shows the residual maps for \texttt{models 2} and \texttt{3}, where the extended emission from the Ancora SNR allows a much better fit to the data with lower residuals.
Note that Fig.~\ref{fig:signmap_compare}~(bottom left and bottom right, respectively) is the same model as Fig.~\ref{fig:sign_map} (left and right, respectively) but with the extended emission from Ancora removed.

\subsection{Multi-wavelength modelling}

The detection of Ancora SNR at radio wavelengths prompted the analysis of the \g\ region. The radio spectral points are thus taken from~\citet{Filipovic_2023}.
At X-ray energies \cite{Filipovic_2023} notes a non-conclusive result by \textit{eROSITA}, as emission is detected uniformly from the whole region. We treat this result as a non-detection. None of our models is predicting any detectable non-thermal synchrotron radiation in the energy band that is observed by \textit{eROSITA} \citep{2012arXiv1209.3114M}.
The \textit{Fermi}--LAT spectral points are taken from the best-fit model using the \pl\ spectral model with and without modelling J1028 (\texttt{model~2 and 3}).

We investigated a leptonic inverse Compton (IC) and a hadronic Pion decay (PD) scenario for the $\gamma$-ray emission and performed \textit{Naima} modelling~\citep{Naima_2015} assuming a power-law distribution of electrons with an exponential cutoff, or a power-law distribution for protons. In the later case, a cutoff cannot reliably be fit due to the lower number of significant data points.

\begin{figure*}
    \centering
    \captionsetup[subfigure]{labelformat=empty}
    \subfloat[\centering ]{{\includegraphics[width=0.86\columnwidth, trim={0.2cm 0cm 0.45cm 0cm}, clip]{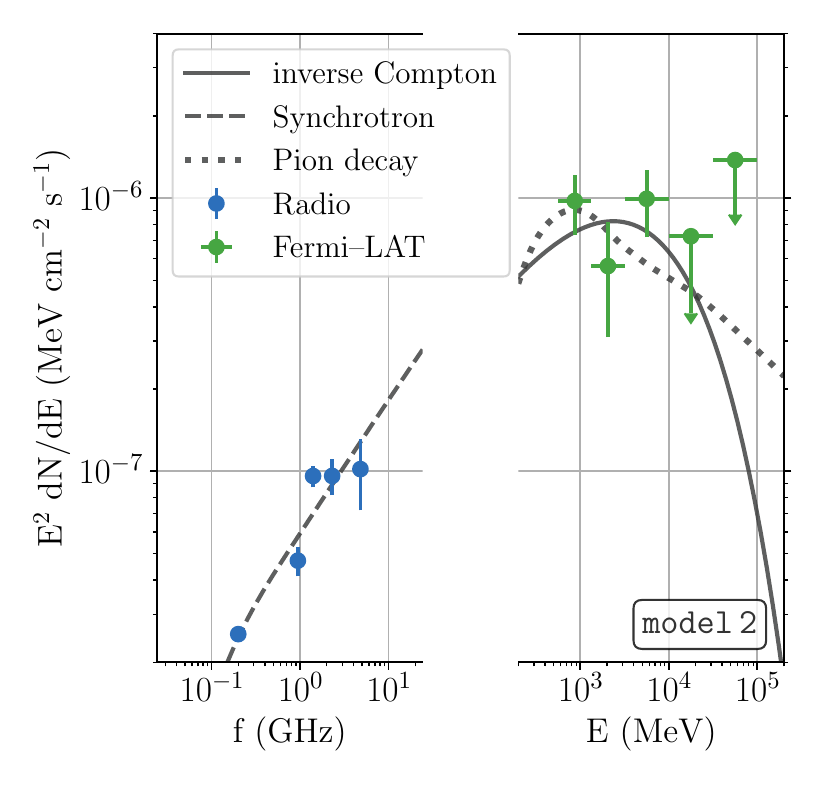}}}%
    \qquad
    \quad
    \qquad
    \subfloat[\centering ]{{\includegraphics[width=0.86\columnwidth, trim={0.1cm 0cm 0.55cm 0cm}, clip]{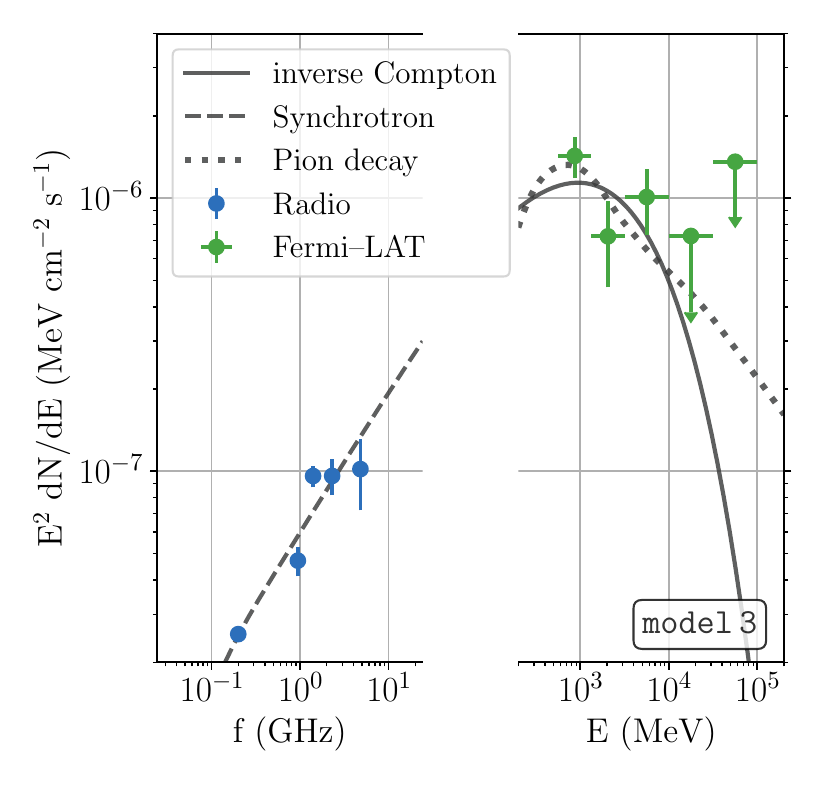}}}

    \vspace{-0.6cm}
    \caption{Multi-wavelength SED from \g. Blue points are radio-flux measurements from \citet{Filipovic_2023} and the \textit{Fermi}--LAT data points represent \texttt{model\,2} (left) and \texttt{model\,3} (right). The curves represent the best-fit \textit{Naima} models with the particle population parameters indicated in Table~\ref{tab:naima}. Solid curves represent the synchrotron emission, dashed curves the inverse-Compton emission and dotted curves Pion-decay emission.}
    \label{fig:Naima_model}
\end{figure*}

The fit parameters for the underlying particle populations are presented in Table~\ref{tab:naima} and the resulting emission spectra in Fig.~\ref{fig:Naima_model}. Note that the hadronic models do not constrain the electron populations and that hence the values of the magnetic field strength $\text{B}_0$ and total energy in electrons $\text{W}_\text{e}$ for the corresponding leptonic models represent lower and upper limits, respectively. We used $\text{n}=\SI{0.15}{\per\cubic\cm}$ as the density of the target material in the hadronic model, taken from within the range of $\SI{0.11}{} -\SI{0.24}{\per\cubic\cm}$, as estimated by \cite{Filipovic_2023} using~\cite{Wolfire_2003}.

The fit parameters for source \texttt{model\,2} and \texttt{model\,3} are very similar and overlap within their respective uncertainties, so we will discuss them together.

We find that the energy in electrons is approximately \SI{0.25}{\percent} of the initial, assumed explosion energy of \SI{e51}{\erg} -- these are typical values expected from a source like the one presented. The spectral index $\text{s}\approx2$, which is mainly determined by signal in the radio bands, matches the canonical expectation for diffusive shock acceleration (DSA).

The cutoff is only weakly constrained. Because of the limited photon counts, we are not able to distinguish between a spectral break and a cutoff. To draw any definite conclusions observations at very-high $\gamma$-ray energies (VHE) are needed. The magnetic field is found to be lower than estimates by \cite{Filipovic_2023}. However, it is roughly comparable with the shock-compressed weak interstellar medium (ISM) field of $\text{B}_0 \leq \SI{3}{\micro G}$ as an off-plane source. 

Due to the low gas density inferred from analysis of the radio observations~\citep{Filipovic_2023}, Pion decay models require a rather large energy budget in CRs of \SI{30}{\percent} of the explosion energy. Taken at face value, this makes a leptonic scenario favourable based on the required energy in electrons and protons, respectively. Additionally, it was found in earlier studies that leptonic remnants in low-density environments tend to reach a higher peak luminosity in $\gamma$-rays~\citep{Brose_2021}. Only the cosmic microwave background (CMB) was considered as the contributing photon field as the object is an off-plane remnant. However, the energy in protons scales linearly with the ambient density, so that a higher ambient density could resolve this tension for the hadronic model. \textit{MOPRA} observations show an increasing density towards Ancora \citep{2018PASA...35...29B} but the remnant itself is not covered in the observations. Contrary, a density of $0.4\,\text{cm}^{-3}$ is close to the environment of Tycho, which is also inferred to be farther away, where clear signatures of thermal X-ray emission are detected. The \textit{eROSITA} image of Ancora in \cite{Filipovic_2023} shows no sign of enhanced emission from the remnant. Additional investigations of the X-ray emission and gas density around Ancora are clearly needed to resolve this issue.

\begin{figure*}
    \centering
    \captionsetup[subfigure]{labelformat=empty}
    \subfloat[\centering ]{{\includegraphics[width=0.85\columnwidth, trim={1.2cm 0cm 3.5cm 0cm}, clip]{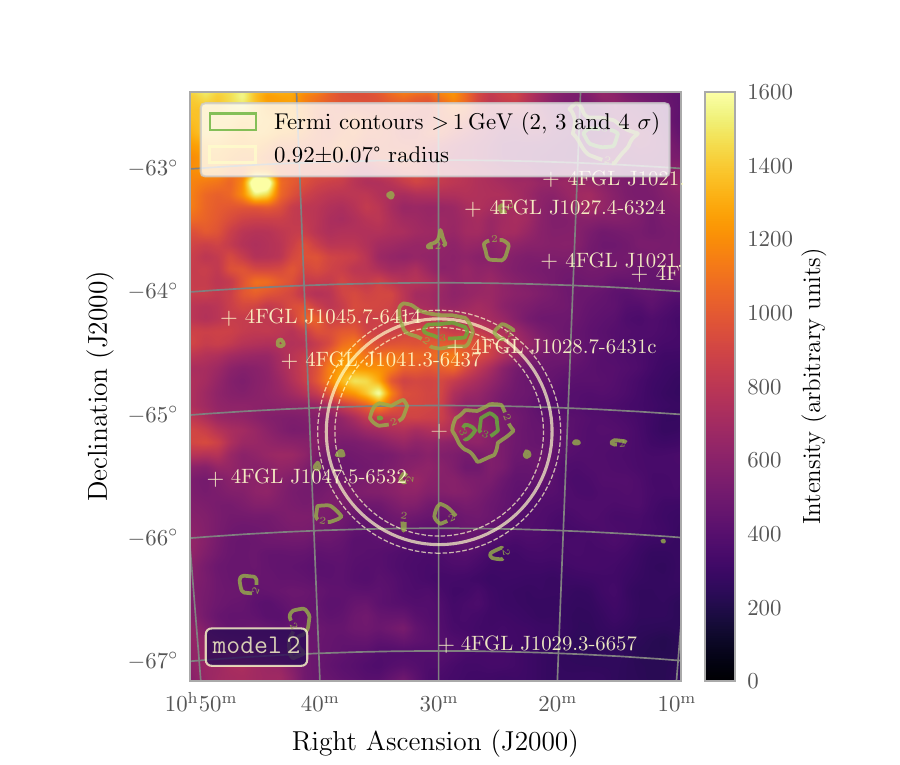}}}%
    \qquad
    \subfloat[\centering ]{{\includegraphics[width=0.965\columnwidth, trim={2.2cm 0cm 1.0cm 0cm}, clip]{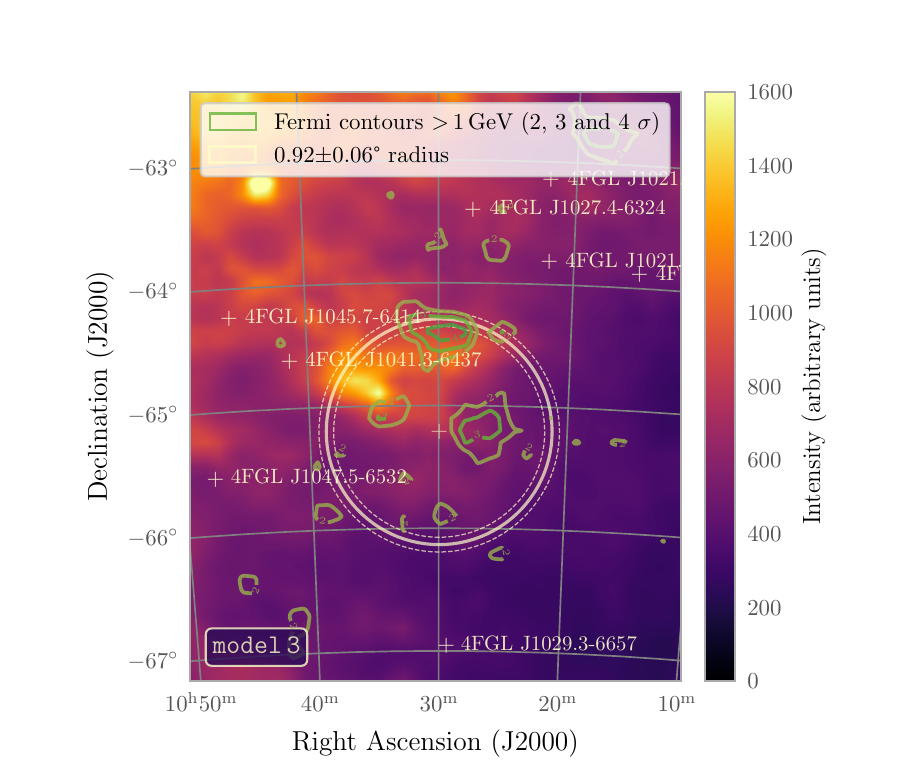}}}%
    \vspace{-0.6cm}
    \caption{Thermal, unpolarised dust emission map produced by the \textit{Planck} mission~\citep{Planck2016}, overlaid with \textit{Fermi}--LAT contours above \SI{1}{\GeV} of Ancora SNR (\texttt{model\,2} (left) and \texttt{model\,3} (right)) and their respective spatial extension.}
    \label{fig:dust_map}
\end{figure*}

\begin{table}
    \renewcommand*{\arraystretch}{1.1}
 	\centering
 	\caption{\textit{Naima} fit parameters. We used an exponentially cutoff power-law model for the underlying electron and proton distributions to fit the data points of \texttt{model\,2} and \texttt{model\,3}.}
 	\label{tab:naima}
 	\begin{tabular}{lllll} 
 		\hline
        \hline
        Model& W$_{\text{e}/\text{p}}$ & s & E$_\text{cut}$ & B$_0$\\
        &(\SI{}{\erg}) & --- & (\SI{}{\GeV}) & (\SI{}{\micro G})\\
        & $>$\SI{1}{\GeV} \\
 		\hline\noalign{\vskip 0.03cm}
        \texttt{2}, leptonic & $1.6^{+0.6}_{-0.4}\times10^{48}$ & $2.0^{+0.1}_{-0.1}$ & $ 700^{+380}_{-260}$ & $6.4^{+1.9}_{-1.4}$\\  [0.03cm]
        \texttt{2}, hadronic & $2.5^{+0.6}_{-0.5}\times10^{50}$ & $2.4^{+0.3}_{-0.2}$ & --- & --- \\  [0.03cm]
        
        \hdashline[0.75pt/1pt]\noalign{\vskip 0.03cm}
        \texttt{3}, leptonic & $2.9^{+1.5}_{-0.9}\times10^{48}$ & $1.93^{+0.07}_{-0.08}$ & $ 400^{+200}_{-130}$ & $4.4^{+1.5}_{-1.3}$\\ [0.03cm]
        \texttt{3}, hadronic & $3.5^{+0.9}_{-0.7}\times10^{50}$ & $2.50^{+0.23}_{-0.14}$ & --- & ---\\[0.05cm]
        \hline
        \hline
 	\end{tabular}
\end{table}

\begin{table*}
    \renewcommand*{\arraystretch}{1.1}
 	\centering
    
 	\caption{Comparison of Ancora SNR to fluxes and photon spectral indices of other known high-latitude SNRs detected at high energies, sorted by their total energy flux. List partially compiled making use of the SNR catalogue provided by \cite{SNRcat_2012}.$^\diamond$}
 	\label{tab:highlat_comparison}
    \begin{threeparttable}
 	\begin{tabular}{lllll} 
 		\hline
        \hline
        Source Name & Extension & \multicolumn{1}{l}{Energy Flux} & Photon Spectral Index & Reference\\
         & (deg) & \multicolumn{1}{l}{(\SI{}{\mega\electronvolt\per\square\cm\per\s})} & --- & \\
         &  & \SI{1}{GeV} -- \SI{1}{\TeV} & \\
        \hline
        Ancora SNR / \g & \SI{0.92}{} & \SI{3.29\pm0.78e-6}{}\tnote{$\bot$} &  $2.31\pm0.11$\tnote{$\bot$} & This work \\
        \hdashline[0.75pt/1pt]
        \hdashline[0.75pt/1pt]
        
        G150+4.5 & \SI{1.5}{} & \SI{5.20e-05}{}\tnote{*} & $1.62\pm0.04_{\text{stat}}\pm 0.22_{\text{sys}}$\tnote{$\dag$} & \cite{Devin_2020} \\
        \hdashline[0.75pt/1pt]
        G17.8+16.7 /  FHES\,J1723.5--0501 & \SI{0.73}{} & \makecell[l]{\hfill \\ \SI{1.38\pm0.26e-05}{}\tnote{$\triangledown$}} & \makecell{$1.83\pm0.02_{\text{stat}}\pm 0.05_{\text{sys}}$ \\ $1.97\pm0.08_{\text{stat}}\pm0.06_{\text{sys}}$} & \makecell[l]{\cite{Araya_2021} \\ \cite{Ackermann_2018}}\\
        \hdashline[0.75pt/1pt]
        G296.5+10.0 / FHES\,J1208.7--5229 & \SI{0.7}{} & \makecell[l]{ \SI{8.17e-06}{}\tnote{**} \\ \SI{1.13\pm0.24e-05}{}\tnote{$\triangledown$}} & \makecell[l]{$1.85\pm0.13$ \\ $1.81\pm0.09_{\text{stat}}\pm0.05_{\text{sys}}$ }& \makecell[l]{\cite{Araya_2013} \\ \cite{Ackermann_2018}}\\
        \hdashline[0.75pt/1pt]
        SN 1006 / G327.6+14.6 & 0.1 & \SI{3.63\pm1.62e-06}{}\tnote{\dag\dag} & $1.57\pm0.11$ & \cite{Condon_2017}\\
        \hdashline[0.75pt/1pt]
        Calvera SNR / G118.4+37.0 & \SI{0.53}{}  & \SI{3.06e-06}{}\tnote{$\triangledown\triangledown$} & $1.66\pm0.10_{\text{stat}}\pm 0.03_{\text{sys}}$ & \cite{Araya_2022} \\
        \hdashline[0.75pt/1pt]
        G166+4.3 & $\sim\SI{0.3}{}$ & \SI{2.87e-06}{}\tnote{**} & $2.7\,\,\,\pm0.1$ & \cite{Araya_2013} \\  
        \hline
        \hline
    \end{tabular}
    
    \begin{tablenotes}
        \scriptsize
        \item[$\diamond $] \href{http://snrcat.physics.umanitoba.ca}{snrcat.physics.umanitoba.ca}
        \item[$\bot$] values taken from \texttt{model\,3}, energy range $\SI{1}{\GeV} - \SI{316}{\GeV}$
        \item [*] calculated using data from Table~2 in ~\cite{Devin_2020}, and using results for the radial Gaussian model and log-parabola spectral model
        \item[\dag] log-parabola model, $\alpha$ given in Table, $\beta = 0.07 \pm 0.02_{\text{stat}}\pm 0.02_{\text{sys}} $.
        \item[$\triangledown$] from FITS data provided with~\cite{Ackermann_2018}
        \item[**] calculated using data from Table~2 in~\cite{Araya_2013}
        
        \item[\dag\dag] range is $\SI{1}{\GeV} - \SI{2}{\TeV}$
        \item[$\triangledown\triangledown$] calculated using data from~\cite{Araya_2022}

    \end{tablenotes}

    \end{threeparttable}
\end{table*}

\section{Discussion} 
\label{sec:discussion}

\subsection{Distinguishing between different models describing the gamma-ray excess}
Although \texttt{model\,2} is statistically slightly favoured over \texttt{model\,3} it seems that overall the two are fairly similar in terms of $\Delta\text{AIC}$ and a significant preference for one over the other is not given. There is no conclusive evidence to assume that J1028 is actually a separate source. When modelling J1028 together with an overlapping disk, its significance decreases from the catalogue value of \SI{5.6}{\SIsigma} to \SI{4.4}{\SIsigma}, below the threshold of what is commonly assumed to be significant in the field. Additionally, modelling an extended disk together with J1028 provides a substantial number of additional degrees of freedom to the overall model. The combined flux from the region is significantly improved when only modelling one extended source. The authors of this work therefore tend to favour the simpler scenario with only one extended model (\texttt{model\,3}).

\subsection{Comparison with other high-latitude SNRs}
With Ancora we add another object to the list of known SNRs emitting $\gamma$-rays at \SI{}{GeV} energies located at high Galactic latitudes, with the other known sources off the Galactic plane, such as SN\,1006~\citep{Condon_2017}, G296.5+10.0~\citep{Araya_2013, Ackermann_2018}, and G166.0+4.3~\citep{Araya_2013}.

We have collected the high-energy data from different publications in Table~\ref{tab:highlat_comparison}, where we took the spectral fits quoted and re-calculated the energy flux for the energy range $\SI{1}{\GeV}-\SI{1}{\TeV}$, consistent with \citet{Ackermann_2018}.

The source FHES~J1741.6--391 has previously been associated with an SNR~\citep{Ackermann_2018}, however it is currently not confirmed as such,\footnote{It has been found to have an extension radius of \SI{15}{\arcmin} at radio wavelengths~\citep{Green_2019}, compared to the extension of \SI{1.35}{\degree} in gamma-rays~\citep{Ackermann_2018}. Due to the large angular size difference it is not clear if the two are physically related.} and has thus not been included in Table~\ref{tab:highlat_comparison}.

For Ancora SNR we obtained a photon spectral index of $\Gamma\approx2.3$.
This value lies between results previously found for other high-latitude SNRs such as Calvera SNR~\citep{Arias_2022, Araya_2022} with $\Gamma=1.66$ at one extreme, and G166+4.3 with $\Gamma=2.7$~\citep{Araya_2013} at the other. The location of the aforementioned SNRs far from the Galactic plane and thus in potentially lower-density environments could make leptonic emission scenarios more likely for them. The observed range of photon spectral indices is not constraining in this regard as hard spectral indexes of $\Gamma\approx1.5$ are expected for a canonical DSA-accelerated electron spectrum of $\text{s}=2$; intermediate indexes around $\Gamma\approx2$ for SNRs with substantial cooling and a potential mix of different electron populations \citep{2022ApJ...926..140S}; and soft spectral indices of $\Gamma\approx2.7$ for cases where particles escape a fading accelerator \citep{Brose_2021}.

When comparing the energy fluxes, the lowest flux is $\approx$\,$\SI{2.87e-6}{\MeV\per\cm\squared\per\s}$ for G166.0+4.3 and the highest flux of $\approx\SI{5.20e-5}{\MeV\per\cm\squared\per\s}$ is found for G150+4.5, which also has the largest angular size.
The energy flux of the Ancora SNR puts it on the lower end of the range (see Table~\ref{tab:highlat_comparison}).
Obtaining the physical luminosity and size of these remnants requires a good estimate of their distance which in most cases is not available, but at least from the spectral index it appears that the Ancora SNR provides an `in-between' example connecting the hard- and soft-spectrum sources.
This may be an indication that it has an intermediate age among the detected SNRs.

\subsection{Possibility of interaction with interstellar clouds}
The extended \textit{Fermi}--LAT source J1028 has `c' appended to its catalogue designation, indicating possible association with molecular clouds along the line of sight.
This possibility was investigated using the thermal, unpolarised, dust emission maps produced by the \emph{Planck}\footnote{Based on observations obtained with Planck (\href{http://www.esa.int/Planck}{esa.int/Planck}), an ESA science mission with instruments and contributions directly funded by ESA Member States, NASA, and Canada.} mission \citep{Planck2016}.
The map was downloaded from the NASA/IPAC Infrared Science Archive\footnote{\href{https://irsa.ipac.caltech.edu/data/Planck/release\_2/all-sky-maps/foregrounds.html}{ irsa.ipac.caltech.edu/data/Planck/release\_2/all-sky-maps/foregrounds.html}} and the contours of \textit{Fermi}--LAT emission were overlaid, shown in Fig.~\ref{fig:dust_map}.
A general gradient can be seen, with dust emission increasing strongly towards the Galactic plane (to the North-East).

There is significant dust emission in the region of \g, partially overlapping with the location of J1028, but the lack of clear correlation with the dust disfavours in principle a sizeable hadronic contribution. However more observations with precise imaging could still unveil denser clumps of gas \citep{2012ApJ...746...82F}, provided that the whole region seems to have a higher dust level than the one expected at such high latitude. Based on this map (Fig. \ref{fig:dust_map}) we expect that leptonic emission is a more likely explanation for the \SI{}{\GeV} emission detected from \g.
Near the region of J1028, it could be that the SNR is interacting with dense gas, and hence that some of the emission from this part of the remnant is hadronic, but this can be neither confirmed nor disproved with the data available. Further observations of the gas and dust distribution around \g\ together with a search for thermal X-ray emission are needed to clearly disentangle any hadronic contribution.

\subsection{Interpretation of the high-energy morphology}
The extension of the $\gamma$-ray emission at around \SI{0.9}{\degree} seems to be slightly larger than that detected at radio wavelengths at about \SI{0.8}{\degree}. This is also observed in the Calvera SNR~\citep{Arias_2022, Araya_2022} and G17.8+16.7~\citep{Ackermann_2018, Araya_2021}, and can be explained by the escape of CRs from an evolved SNR ~\citep{Brose_2021}.
The SNR shock compresses the ambient field, giving rise to radio emission through the synchrotron process in the downstream of the shock, whereas escaping electrons can interact with the uniformly distributed ambient photons of the CMB through IC interaction to produce $\gamma$-rays at much larger distances from the SNR shell.  Thus, the radio and $\gamma$-ray morphology can differ quite significantly, as the synchrotron emission tracks only a sub-population of the electrons accelerated in the Ancora SNR. In that sense, the magnetic field derived by our modelling (see Table \ref{tab:naima}) represents strictly a lower limit to the magnetic field strength in the Ancora SNR. A significantly larger extension of \g\ in VHE $\gamma$-rays can be expected in this scenario. However, based on the observational uncertainties of the measured extension, no firm claim can be made at the moment that the $\gamma$-ray emission is larger than the radio extension.

Additionally, further data at very-high energies (VHE, $>$\SI{100}{\GeV}) for these high-latitude SNRs would strongly constrain the emission processes through spectral modelling (see Fig.~\ref{fig:Naima_model}) and close the gap in observational data between the young historical remnants and old SNRs interacting with molecular clouds \citep{2019ApJ...874...50Z}. The typically large extension of these high-latitude SNRs simplifies morphological comparisons between radio, high-energy and VHE emission and provides valuable information to understand CR escape to the ISM.
An extrapolation of the GeV-flux data to TeV energies suggests that the Ancora and other high-latitude SNRs may be detectable with Imaging Atmospheric Cherenkov Telescopes (IACTs) and potentially also Water Cherenkov Detectors.

So far, SNRs with confirmed hadronic emission signatures have only been observed in the Galactic plane and are interacting with molecular clouds in high-density regions~\citep{2013Sci...339..807A,2019ApJ...874...50Z}. It is however expected that signatures of CR escape -- spectral breaks in the $1-\SI{100}{\GeV}$ range and soft spectra at higher energies -- are also present without such special circumstances, and depend on the age of the remnant~\citep{Brose_2021}.
SNRs at high Galactic latitudes such as the Ancora SNR can help to build an unbiased picture of the evolution of particle acceleration and escape in SNRs.

\section{Conclusions}
\label{sec:conclusions}

Prompted by the recent discovery of an SNR at large Galactic latitude with \textit{ASKAP} we reanalysed the \g\ region with the \textit{Fermi}--LAT instrument and detected a highly extended high-energy $\gamma$-ray source coinciding with the radio position. Different spatial and spectral models were tested in the process. It was found that two models are quite closely favoured, both fitting the excess using a radial disk spatial model and a power-law spectral model. The two scenarios (\texttt{model\,2} and \texttt{3}) only differ in whether they include the modelling for J1028, an unidentified Fermi-catalogue source that lies within \SI{0.7}{\degree} of the SNR centre, or not.
Energy flux estimates for \texttt{model\,2} return values of \SI{4.29\pm1.03e-06}{\MeV\per\square\cm\per\s} (photon flux of \SI{2.29\pm0.45e-08}{\photon\per\square\cm\per\s}) and a spectral index $\Gamma=2.21\pm0.12$, and a significance of \SI{6.7}{\SIsigma} while J1028, modelled by a log-parabola spatial model is left with a significance of \SI{4.4}{\SIsigma}, an energy flux of \SI{7.49\pm2.16e-07}{\MeV\per\square\cm\per\s} (photon flux of \SI{9.68\pm3.08e-10}{\photon\per\square\cm\per\s}), resulting in a total energy flux of $\sim\SI{5.04e-6}{\MeV\per\square\cm\per\s}$ with a total significance of \SI{8.0}{\SIsigma}, if we assume that the two sources combined can both be attested to emission from the SNR.
Considering \texttt{model\,3}, we measured an energy flux of \SI{4.80\pm0.91e-06}{\MeV\per\square\cm\per\s} (photon flux of \SI{3.14\pm0.41e-08}{\photon\per\square\cm\per\s}) and a spectral index $\Gamma=2.32\pm0.11$.
In both cases, the spectrum extends up to around \SI{5}{\GeV}. Morphological hotspots above \SI{1}{\GeV} are well-correlated with the bright western part of the radio shell. 
The best-fit extension was found to be $\sim\SI{0.92}{\degree}$ -- slightly larger than the radio shell at around \SI{0.8}{\degree} -- as was expected from theory.
Given the estimated gas density in the region, the estimated distance and age of the SNR and total energy arguments, the emission is more likely to be of leptonic origin, but a hadronic scenario cannot be ruled out based on the incomplete picture of the gas-distribution around Ancora.

The observations were put in a multi-wavelength context by using \textit{Naima} modelling. We found that either a electron distribution with an exponential cutoff at $\text{E}_\text{max}=(700^{+380}_{-260})\,$\SI{}{\GeV} ($\text{E}_\text{max}=(400^{+200}_{-130})\,$\SI{}{\GeV}), a spectral index of $s=2.0\pm0.1$ ($s=1.93\pm0.07$) and a total energy of $\text{W}_\text{e}=1.6^{+0.6}_{-0.4}\times 10^{48}$\,\SI{}{\erg} ($\text{W}_\text{e}=2.9^{+1.5}_{-0.9}\times 10^{48}$\,\SI{}{\erg}) in electrons above \SI{1}{\GeV} describes the emission well when the magnetic-field strength is $\text{B}_0=6.4^{+1.9}_{-1.4}$\,\SI{}{\micro G} ($\text{B}_0=4.4^{+1.5}_{-1.3}$\,\SI{}{\micro G}) for our models including (excluding) J1028. A hadronic scenario with $\text{s}=2.4^{+0.3}_{-0.2}$ ($\text{s}=2.50^{+0.23}_{-0.14}$) and a total energy of $\text{W}_\text{p}=2.6^{+1.1}_{-0.6}\times10^{50}$\,\SI{}{\erg} ($\text{W}_\text{p}=3.5^{+0.9}_{-0.7}\times10^{50}$\,\SI{}{\erg}) in protons yields a similarly good fit. The cutoff energy, however, cannot be constrained in that case.

Ancora is only the seventh confirmed SNR detected at high Galactic latitude with \textit{Fermi}--LAT. This new population of remnants has the potential to constrain the physics of particle diffusion and escape from SNRs into the Galaxy.
Due to their narrow field of view of a few degrees, IACTs usually only observe off the Galactic plane if clear indications of emission from these regions are given. Extrapolating the energy flux and considering the brightness of the Ancora SNR, this object may be observable with current IACTs such as \textit{H.E.S.S.}, and should be easily detectable with next-generation IACTs such as the \textit{Cherenkov Telescope Array~(CTA)}, and possibly future Water Cherenkov detectors such as the \textit{Southern Wide-field Gamma-ray Observatory (SWGO)}.


\section*{Acknowledgements}
CBS acknowledges support from a Royal Society-Science Foundation Ireland Research Fellows Enhancement Award 2021 (22/RS-EA/3810).
This work was supported by an Irish Research Council (IRC) Starting Laureate Award (IRCLA\textbackslash 2017\textbackslash 83).
RB acknowledges funding from the Irish Research Council under the Government of Ireland Postdoctoral Fellowship program.
JM acknowledges support from a Royal Society -- Science Foundation Ireland University Research Fellowship (20/RS-URF-R/3712)
MDF, GR and SL acknowledge Australian Research Council funding through grant DP200100784. IS acknowledges support by the National Research Foundation of South Africa (Grant Number 132276).

Special thanks to M.~Lemoine-Goumard, J.~Devin and Q.~Remy for fruitful discussions. We also want to thank the anonymous reviewer for helpful comments improving the manuscript.

\vspace{0.1cm}
CBS would like to dedicate this work to Daniel~R.~W.~Bale, who, through his creative and unconventional yet highly systematic approach to experiments and projects in various fields, paired with his profound curiosity, shaped the lives of many whose paths he crossed.


\section*{Data Availability}
This work makes use of publicly available \textit{Fermi}--LAT data provided online by the Fermi Science Support Center at
\href{http://fermi.gsfc.nasa.gov/ssc/}{fermi.gsfc.nasa.gov/ssc/}, as well as publicly available data obtained with \textit{Planck} (\href{http://www.esa.int/Planck}{esa.int/Planck}), an ESA science mission.

This research made use of \textit{Fermitools}~(\href{https://github.com/fermi-lat/Fermitools-conda}{github.com/fermi-lat/Fermitools-conda}), as well as the following Python packages: \textit{Fermipy}~\citep{Atwood_2013}, \textit{Astropy}~\citep{Astropy_2022}, \textit{Numpy}~\citep{Numpy_2020}, and \textit{Matplotlib}~\citep{Hunter_2007}.

Access to the scripts used to produce this work will be made available upon contact with the first author.




\bibliographystyle{aa}
\bibliography{bibliography} 




\appendix


\section{Naming of Ancora SNR}\label{appendix:ancora}

\begin{figure}[h!]
    \begin{center}
	\includegraphics[width=0.85\columnwidth, trim={18.0cm 2cm 21cm 10cm}, clip]{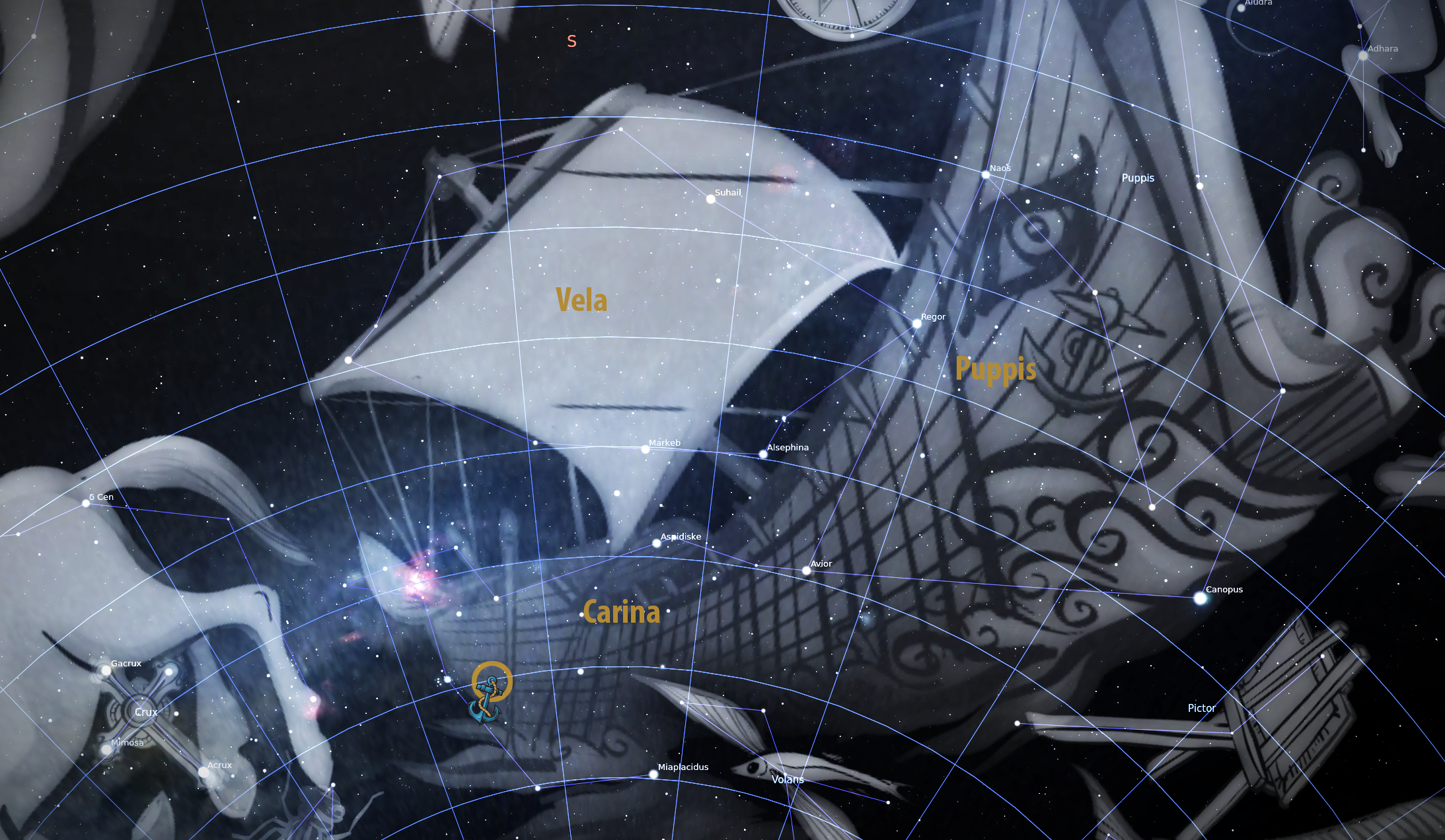}
    \end{center}
    	\caption{\textit{Ancora} in the context of \textit{Argo navis}, marked with a golden circle around the \g\ position. Graphics: Stellarium \citep{Stellarium_2023} and \textit{\href{https://www.pngegg.com}{pngegg}}.}
	\label{fig:argo-navis}
\end{figure}

Historically, \textit{Argo navis} (Engl.: the ship Argo) was one of the 48 constellations of Ptolemy~\citep{Ptolemy_1998}. The name Argo referred to the ship Argo in Greek mythology, with which Jason and his crew, the Argonauts, sailed to Colchis, a place on the shores of the Black Sea, to recover the Golden Fleece~\citep{Ridpath_2018}. In 1930 at the meeting of the International Astronomical Union (IAU) where the decision on settling on the modern 88 constellations was made~\citep{Delporte_1930}, it was officially broken up into three distinct constellations, as had been suggested by Nicolas Louis de Lacaille in his 1755 catalogue~\citep{DeLacaille_1755}: \textit{Puppis}, the poop deck or stern; \textit{Vela}, the sail; and \textit{Carina}, the keel. \textit{Argo navis} was officially retired.

The supernova remnant discussed in this work at GLON/GLAT~$ =\SI{288.8}{\degree}/\SI{-6.3}{\degree}$ is located in the constellation of \textit{Carina}, towards the front of the keel, a position where the anchor of a ship would usually be found. To keep in line with Latin naming traditions, the Latin term for anchor, \textit{ancora}, was henceforth used to distinguish the SNR from others. An artist's impression of the anchor in context can be found in Fig.~\ref{fig:argo-navis}.

\end{document}